# On the transition from photoluminescence to thermal emission and its implication on solar energy conversion

Assaf Manor[1], Leopoldo L. Martin[2] and Carmel Rotschild[1,2*]

[1] *Russell Berrie Nanotechnology Institute, Technion – Israel Institute of Technology, Haifa 32000, Israel*

[2] *Department of Mechanical Engineering, Technion – Israel Institute of Technology, Haifa 32000, Israel*

Email: carmelr@technion.ac.il

Photoluminescence (PL) is a fundamental light-matter interaction, which conventionally involves the absorption of energetic photon, thermalization and the emission of a red-shifted photon. Conversely, in optical-refrigeration the absorption of low energy photon is followed by endothermic-PL of energetic photon. Both aspects were mainly studied where thermal population is far weaker than photonic excitation, obscuring the generalization of PL and thermal emissions. Here we experimentally study endothermic-PL at high temperatures. In accordance with theory, we show how PL photon rate is conserved with temperature increase, while each photon is blue shifted. Further rise in temperature leads to an abrupt transition to thermal emission where the photon rate increases sharply. We also show how endothermic-PL generates orders of magnitude more energetic photons than thermal emission at similar temperatures. Relying on these observations, we propose and theoretically study thermally enhanced PL (TEPL) for highly efficient solar-energy conversion, with thermodynamic efficiency limit of 70%.

**Main text:**

Endothermic photoluminescence has mainly been studied in the framework of optical refrigeration where a narrow-line pump at the absorption tail of a photoluminescent material leads to endothermic PL at shorter wavelengths, thereby extracting heat[1]. Recent studies have demonstrated cryogenic temperatures with quantum efficiency (QE) approaching unity[2]. At such low temperatures, thermal population is negligible with respect to the PL excitation. In contrast, at high temperatures, the PL and thermal populations compete for dominance. In general, PL requires directional energy transfer from the excited to the emitting modes. This only occurs if the excitation rate is above the rate of thermal excitation where directional energy transfer is canceled by thermal equilibrium. This tradeoff is expressed by the generalized Planck's law, suggested by Würfel (equation (1)), describing the spontaneous emission rate of a bandgap material [3,4]:

$$R(\hbar\omega, T, \mu) = \varepsilon(\hbar\omega) \cdot \frac{(\hbar\omega)^2}{4\pi^2 \hbar^3 c^2} \frac{1}{e^{\frac{\hbar\omega - \mu}{KT}} - 1} \cong \varepsilon(\hbar\omega) \cdot R_0(\hbar\omega, T) \cdot e^{\frac{\mu}{K_b T}} \qquad (1)$$

Where $R$ is the emitted photon flux (photons per second per unit area), $T$ is the temperature, $\mu$ is the chemical potential, $\varepsilon$ is the emissivity, $\hbar\omega$ is the photon energy, and $K_b$ is Boltzmann's constant. The chemical-potential ($\mu>0$) defines the deviation of the excitation from thermal equilibrium. The chemical potential $\mu$ is frequency-invariant as long as thermalization equalizes excitation between modes. This is true for excited electrons in the conduction band of solid-state semiconductors as well as for excited electrons in isolated molecules[3]. By definition, the regime where PL and thermal excitations compete for dominance is when $\mu$ approaches zero. At $\mu=0$, the radiation is reduced to the thermal emission rate ($R_0$).

We start with simulating the emission of an ideal bandgap material, chosen to be $E_g=1.3\ eV$, with unity absorption of photons above the bandgap, zero absorption of sub-bandgap photons and unity QE. This is done by balancing the incoming and outgoing rates at the PL absorber for both the energy and photon fluxes, separately. We keep the absorbed photon rate constant, while increasing the absorbed energy rate. This simulates the increase in energy of each absorbed photon, thereby increasing heat dissipation at constant photon rate. Once the incoming photon and energy rates are set, the thermodynamic quantities $T$, $\mu$, and the emission spectrum are uniquely defined. Intuitively, the only way to conserve these two rates is if each emitted photon is blue-shifted, thereby extracting heat. It is constructive to present the radiation as function of temperature, as shown in figure 1a. As evident, at low temperatures, the emission's line shape at the band-edge is narrow, and is blue-shifted with temperature increase. The red portion of the emission represents the thermal population, $R_0$, which increases and becomes visible at high temperatures. The temperature increase leads to the reduction in the chemical potential, $\mu$, as depicted in the inset of figure 1a. This trend continues until $\mu=0$, where the emission becomes purely thermal, and the balancing of photon flux does not apply. The transition to thermal emission, which is completely disordered, results in a sharp increase of the photon rate to its maximally possible value for given energy.[5]

We also compare the ability of PL and thermal emission to extract heat by emitting energetic photons. Figure 1b compares the emitted rate of photons with energy above *1.45 eV* (corresponding to *λ<850nm*) and the total emitted photon rate (1b inset), for two materials at various temperatures. The first is PL material, with bandgap at *1.3 eV* as simulated previously (blue line), while the second is a black-body with step function emissivity at *1.3eV* (red line). As can be seen, the emitted rate of energetic photons in the PL material is orders of magnitude greater than in thermal emission, as long as long as $\mu>0$, while the total photon rate is conserved (inset). This indicates that the increase in emitted energy with temperature is due to the blue shift of each emitted photon. At $\mu=0$ the two emissions converge and the photon rate abruptly rises.

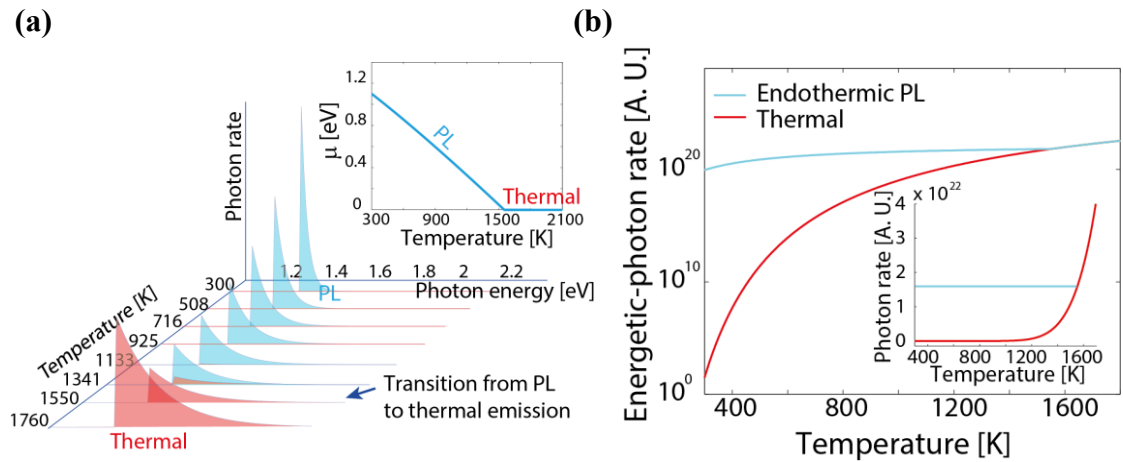

**Fig. 1. Theoretical analysis and results** (a) Emission evolution of PL material with temperature. (b) Emission rates of energetic photons and total photons rate (inset) for PL (blue line) and thermal emission (red line) at various temperatures.

Experimentally, the need for high QE PL at high temperatures limits the use of solid-state semiconductors, due to reduction in their QE by temperature-dependent non-radiative recombination mechanisms[6]. One the other hand, rare-earth ions such as Neodymium and Ytterbium are excellent choice of materials as their electrons are localized and insulated from interactions. This results in the conservation of their high QE at extremely high temperatures[7]. Due to the lack of interaction between electrons, each energy gap can be populated at different $\mu$ values, but thermalization equalizes $\mu$ within each spectral band.

Specifically, we experimentally study the transition between the PL and thermal emission at the 905nm fluorescence band of Neodymium ($Nd^{+3}$) doped silica fiber tip under 532nm PL excitation. The 905nm emission corresponds to the transition between the $^4F_{3/2}$ and $^4I_{9/2}$ energy levels, the latter being the ground state, which is essential for maximizing the thermal radiation signal. In order to control the heat current separately from the PL excitation, we use a $CO_2$ laser operating at *10.6μ* wavelength. At this wavelength the photons are efficiently absorbed by the silica[8], and converted to a constant heat flow. The temperature and the chemical potential are uniquely defined by these two currents.

The experimental setup is sketched in figure 2a. The power spectrum is measured by a calibrated spectrometer. A weak PL excitation at 532nm is kept constant at *1mW*, while the $CO_2$ laser power varies between 0 to *150mW*. We monitor the temperature by Fluorescence Intensity Ratio Thermometry (FIR)[9] (see supplementary information). The spectral results are shown in figure 2b. As we increase the thermal load, the temperature increases, and the PL exhibits a blue-shift evolution. This is shown by the reduction in the emission of the low photon-energy peak at 905nm, and enhancement of the high photon-energy peak at 820nm. This trend continues until it reaches the transition temperature of *~1500K*. As we increase the temperature further, the number of emitted photons increases sharply at all wavelengths (dotted lines in Figure 2b). Figure 2c shows the total number of emitted photons at wavelengths between 600nm and 1000nm under PL excitation at various temperatures (blue line). In order to compare to thermal emission under equivalent conditions, we turn off the weak PL pump while the thermal current is unchanged (red line). The thermal emission at temperatures lower than *1150K* is below our detection limit and we extrapolate the experimental values down to *300K*. The inset of figure 2c shows that the extracted chemical potential values (blue dots) are in good agreement with the theory (grey line) (see detailed description at the supplementary information).

As demonstrated, the total photon rate is conserved at various temperatures, as long as $\mu>0$, and increases sharply at $\mu=0$, while converging with the thermal emission.

In Figure 2d we compare the emission rate of energetic photons at wavelengths between 600nm-850nm under PL excitation of *1mW* (blue line) and *4mW* (purple line), to thermal emission (red line). As can be seen, at temperatures below *1300K* the energetic-photon rates under PL excitation exceed that of thermal emission at similar temperatures by orders of magnitude.

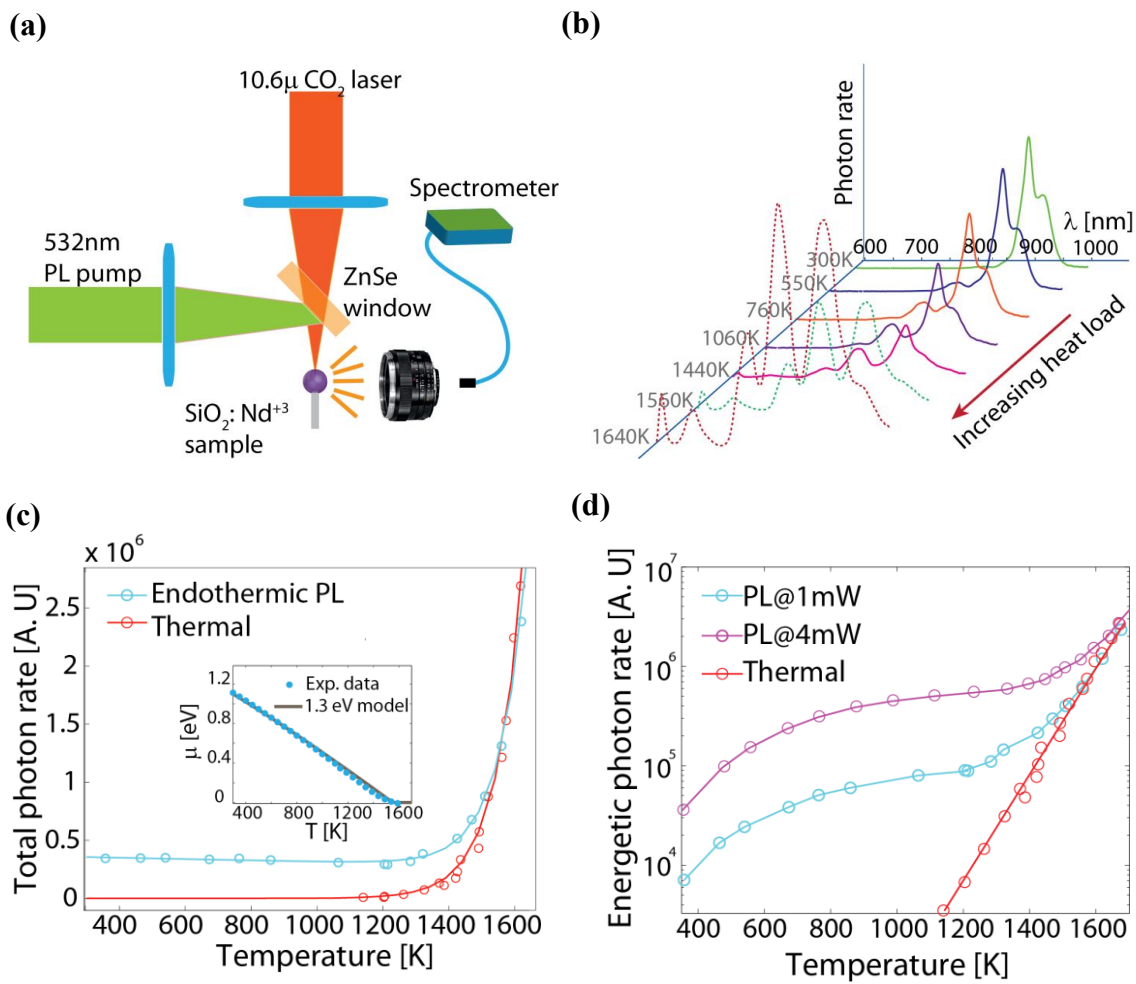

**Fig. 2. $Nd^{+3}$ experimental results** (a) Experimental setup (b) PL spectra evolution with temperature (c) Total photon rate of PL and thermal emissions (inset: chemical potential Vs. temperature) (d) The energetic photon ($\lambda<850nm$) rate for PL and thermal emissions.

The efficient conversion of heat to high-energy photons by endothermic PL operating at comparably low temperatures opens the way for practical harvesting of thermal energy. In Photovoltaics (PV), The Shockley-Queisser (SQ) efficiency limit for single-junction solar-cells[10] is to a great extent due to inherent heat dissipation accompanying the quantum process of electro-chemical potential generation. Concepts such as solar thermo-photovoltaics[11–13] (STPV) and thermo-photonics[14] aim to harness this dissipated heat, yet exceeding the SQ limit has not been achieved, mainly due to the requirement to operate at very high temperatures. In the following section, we propose and analyze Thermally Enhanced PL (TEPL) device for solar energy conversion.

Figure 3a shows the conceptual design. A thermally-insulated low-bandgap photoluminescent absorber completely absorbs the solar spectrum above its bandgap and emits endothermic PL towards a higher bandgap PV cell, maintained at room temperature. The PV converts the excessive thermal energy to electricity. For minimizing radiation losses, a semi-ellipsoidal reflective dome recycles photons by reflecting back emission at angles larger than the solid angle $\Omega_l$[15,16]. In addition, the PV's back-reflector[17,18] reflects sub-bandgap photons back to the absorber. Figure 3b shows the absorber and PV energy levels, the endothermic PL and the current flow, as well as the reflected photons from the PV. As before, the absorber's thermodynamics is governed by two conservation rules: **i.** conservation of energy. **ii.** balancing of photon flux, which takes into account the absorbed solar photons and spontaneous emission of both the absorber and the PV. At the PV, only balancing of photon flux is applied between the incoming PL, the PV's emitted light and the extracted electric current, $R_{eh}$. Energy conservation does not apply, because dissipated heat ($E_{dissipated}$) is removed from the PV to keep it at room temperature. The different fluxes are depicted in figure 3c.

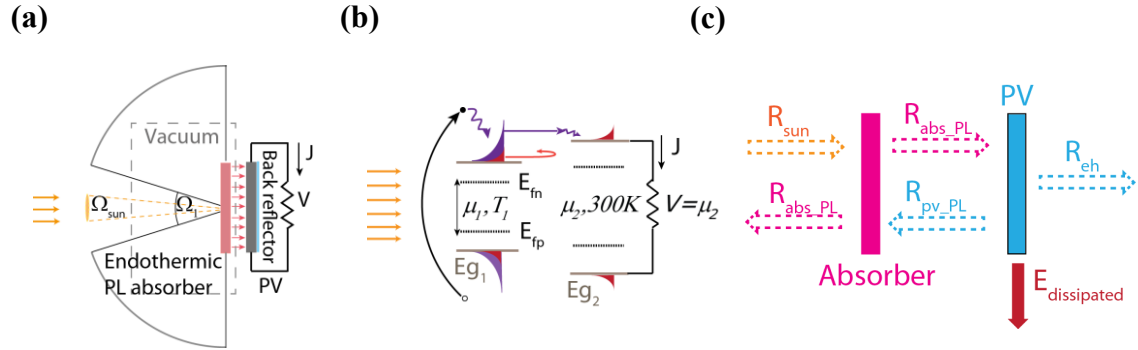

**Fig. 3. Endothermic PL based device for PV** (a) The device's scheme (b) The absorber and PV energy levels with the currents flow. (c) Rates included in the detailed balance (dotted arrows) and heat removal (solid red arrow)

Assuming unity QE for both the emitter and the PV, the detailed-balance of photon and energy fluxes is uniquely accommodated by fixing the only free parameter in the system: the PV voltage. This defines the absorber's thermodynamic properties, $T$ and $\mu$, as well as the device's *I-V* curve. A rigorous analysis of the detailed balance is given in the supplementary information. Theoretical maximal efficiency requires photon recycling at all angles except the acceptance solid angle of the sun ($\Omega_1=\Omega_{sun}=6.94 \cdot 10^{-5}$ *Srad*). However, to be more realistic we analyze the case of non-ideal photon recycling, where portion of the hemisphere is open to the sky; $\Omega_1= 10\Omega_{sun}$. Figure 4a depicts three different *I-V* curves calculated for $Eg_1=0.7$ *eV*, while $Eg_2$ varies between *0.7 eV, 1.1 eV*, and *1.5 eV*. The *I-V* curve for $Eg_1= Eg_2=0.7$ *eV* (red line) shows a remarkable, "double humped" feature, and includes a thermal and a PL Maximal Power Points (MPP). While at the thermal-MPP, enhanced current leads to 53.6% efficiency, the PL-MPP remains at the SQ efficiency limit of 33% (under equivalent photon recycling). When $Eg_2$ is increased, the current enhancement vanishes and is replaced by voltage enhancement, which for $Eg_2= 1.5eV$ sets an efficiency of 65.6% (blue curve), at working temperature of 1300K. The transition between the thermal and the PL MPP's is characterized by a sharp decrease in temperature and increase in the chemical potential as elaborated at the supplementary information.

Figures 4b and 4c show the efficiency and temperature maps at the PL-MPP for various bandgap configurations assuming ideal photon recycling. For $Eg_1=Eg_2$, the efficiency is identical to the maximal SQ limit, and maximal efficiency of 70% is reached for *$Eg_1=0.5eV$* and *$Eg_2=1.4eV$* at *T=1180K*. Although the theoretical efficiencies are similar to that of STPV, operating at temperatures of *2000K-2500K*[19,20], the TEPL operating temperatures are within the range of *1000K-1400K*. Matching the expected maximal efficiency for the experimental parameters where the absorber is $Nd^{+3}$ with energy-gap at *1.3eV* coupled to GaAs PV with bandgap at *1.45eV* (cut off wavelength at 850nm) results in 45% at temperature of *740K*. This assumes optimal sensitization of the $Nd^{+3}$ in order to absorb all the solar radiation at *λ<900nm*.

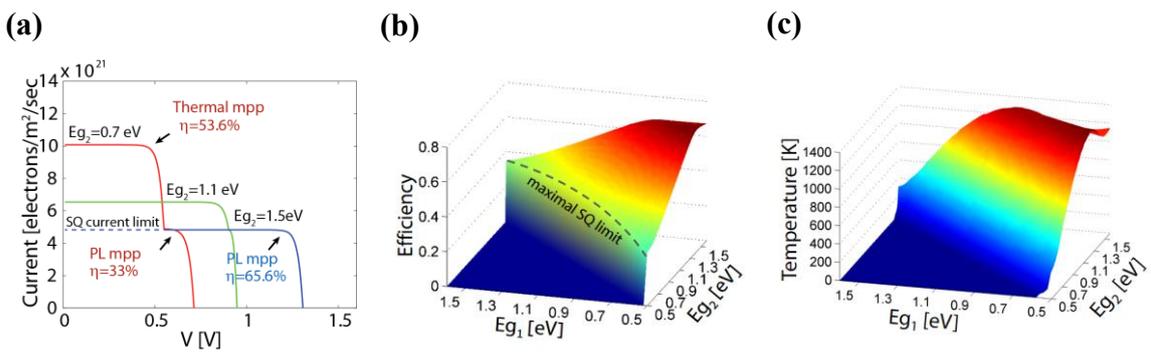

**Fig. 4. Theoretical device analysis** (a) *I-V* curves corresponding to a *0.7 eV* absorber and PV bandgap values of (*0.7 eV, 1.1 eV* and *1.5 eV*) (b) Efficiency and (c) temperature maps, at various bandgap configurations.

From engineering considerations, it is important to arrest sub-bandgap radiative losses through phononic excitations. Such Infrared radiation can be minimized by choosing materials with low emissivity within the relevant spectral range, depending on the operating temperature. For operating at temperatures between *1200K* and *1500K* the peak radiative losses for emissive material are at *2.4μm-1.9μm* respectively. By Kirchhoff's law of thermal radiation low emissivity materials are transparent. Therefore, matrix materials such as Silica, that are transparent at these wavelengths, are preferable choice.

To conclude, by investigating PL in the high-temperature regime, where the thermal population competes with the photonic excitation for dominance, we have demonstrated the generalization of PL and thermal emissions. We have experimentally showed that PL conserves the photon rate with temperature increase, while each photon is blue shifted. We also showed how at the critical temperature, where the chemical potential of radiation vanishes, the PL abruptly transforms into thermal emission and the photon rate increases sharply. In our final experimental demonstration we showed how endothermic-PL generates orders of magnitude more energetic photons than thermal emission at similar temperatures. We take advantage of this feature in order to suggest and analyze a new type of solar energy converter, namely, the Thermally-Enhanced PL device. This device can in theory reach ultra-high efficiencies, in lowered operating temperatures than existing STPV concepts, which may pave the way for disruptive technology in photovoltaics.

**References:**

1. Sheik-Bahae, M. & Epstein, R. I. Laser cooling of solids. *Laser Photonics Rev.* **3,** 67–84 (2009).

2. Seletskiy, D. V. *et al.* Laser cooling of solids to cryogenic temperatures. *Nat. Photonics* **4,** 161–164 (2010).

3. Wurfel, P. The chemical potential of radiation. *J. Phys. C Solid State Phys.* **15,** 3967–3985 (1982).

4. Herrmann, F. & Würfel, P. Light with nonzero chemical potential. *Am. J. Phys.* **73,** 717 (2005).

5. Weinstein, M. A. Thermodynamic limitation on the conversion of heat into light. *J. Opt. Soc. Am.* **50,** 597 (1960).

6. Sze, S. M. & Ng, K. K. *Physics of Semiconductor Devices*. (Wiley-Interscience, 2006).

7. Licciulli, A. *et al.* The challenge of high-performance selective emitters for thermophotovoltaic applications. *Semicond. Sci. Technol.* **18,** S174–S183 (2003).

8. McLachlan, A. D. & Meyer, F. P. Temperature dependence of the extinction coefficient of fused silica for $CO_2$ laser wavelengths. *Appl. Opt.* **26,** 1728–1731 (1987).

9. Berthou, H. & Jörgensen, C. K. Optical-fiber temperature sensor based on upconversion-excited fluorescence. *Opt. Lett.* **15,** 1100–1102 (1990).

10. Shockley, W. & Queisser, H. J. Detailed Balance Limit of Efficiency of p-n Junction Solar Cells. *J. Appl. Phys.* **32,** 510 (1961).

11. Wurfel, P. & Ruppel, W. Upper limit of thermophotovoltaic solar-energy conversion. *IEEE Trans. Electron Devices* **27,** 745–750 (1980).

12. Swanson, R. M. A proposed thermophotovoltaic solar energy conversion system. *IEEE Proc.* **67,** 446 (1979).

13. Lenert, A. *et al.* A nanophotonic solar thermophotovoltaic device. *Nat. Nanotechnol.* **9,** 126–130 (2014).

14. Harder, N.-P. & Green, M. A. Thermophotonics. *Semicond. Sci. Technol.* **18,** S270 (2003).

15. Kosten, E. D., Atwater, J. H., Parsons, J., Polman, A. & Atwater, H. A. Highly efficient GaAs solar cells by limiting light emission angle. *Light Sci. Appl.* **2,** e45 (2013).

16. Braun, A., Katz, E. A., Feuermann, D., Kayes, B. M. & Gordon, J. M. Photovoltaic performance enhancement by external recycling of photon emission. *Energy Environ. Sci.* **6,** 1499–1503 (2013).

17. Yablonovitch, E., Miller, O. D. & Kurtz, S. R. The opto-electronic physics that broke the efficiency limit in solar cells. in *2012 38th IEEE Photovoltaic Specialists Conference (PVSC)* 001556–001559 (2012).

18. Kayes, B. M. *et al.* 27.6% Conversion efficiency, a new record for single-junction solar cells under 1 sun illumination. in *2011 37th IEEE Photovoltaic Specialists Conference (PVSC)* 000004–000008 (2011).

19. Davies, P. A. & Luque, A. Solar thermophotovoltaics: brief review and a new look. *Sol. Energy Mater. Sol. Cells* **33,** 11–22 (1994).

20. Würfel, P. & Würfel, U. *Physics of Solar Cells: From Basic Principles to Advanced Concepts*. (John Wiley & Sons, 2009).

**Acknowledgment**s

The authors would like to acknowledge Prof. Peter Würfel and Prof. Eli Yablonovitch for their support, fruitful discussions and valuable insights. The authors would also like to thank Asst. Prof. Avi Niv for constructive brainstorming and Dr. Alexander Bekker for help with fabrication of samples.

**Author contributions:**

A. Manor contributed to the theoretical and experimental study, as well as to writing the paper. L. Martin contributed to the FIR experiment. C. Rotschild contributed to the concept and to the paper writing.

**Funding**: This report was partially supported by the by the Russell Berrie Nanotechnology Institute (RBNI) and the Grand Technion Energy Program (GTEP) and is part of The Leona M. and Harry B. Helmsley Charitable Trust reports on Alternative Energy series of the Technion and the Weizmann Institute of Science. We also would like to acknowledge partial support by the Focal Technology Area on Nanophotonics for Detection. A. Manor thanks the Adams Fellowship program for financial support. Prof. C. Rotschild thanks the Marie Curie grant for its support.

**Figure legends:**

**Fig. 1. Theoretical analysis and results** (a) Emission evolution of PL material with temperature. (b) Emission rates of energetic photons and total photons rate (inset) for PL (blue line) and thermal emission (red line) at various temperatures.

**Fig. 2. Nd$^{+3}$ experimental results** (a) Experimental setup (b) PL spectra evolution with temperature (c) Total photon rate of PL and thermal emissions (inset: chemical potential Vs. temperature) (d) The energetic photon ($\lambda$<850nm) rate for PL and thermal emissions.

**Fig. 3. Endothermic PL based device for PV** (a) The device's scheme (b) The absorber and PV energy levels with the currents flow. (c) Rates included in the detailed balance (dotted arrows) and heat removal (solid red arrow)

**Fig. 4. Theoretical device analysis** (a) *I-V* curves corresponding to a *0.7 eV* absorber and PV bandgap values of (*0.7 eV, 1.1 eV* and *1.5 eV*). (b) Efficiency and (c) temperature maps, at various bandgap configurations.

**Fig. S1. FIR experiment** (a) The FIR measurement experimental setup (b) and the intensity ratio vs. temperature plot.

**Fig. S2.** Measured relative emissivity of Nd:SiO$_2$ sample.

**Fig. S3.** The temperature (red) and chemical potential (green) dependence on the PV-voltage for the case of *Eg$_1$=Eg$_2$=0.7eV*.